# Modeling Chromate Removal Using Ion Exchangers in Drinking Water Applications

Fred Ghanem, and Kirti M. Yenkie

Department of Chemical Engineering, Henry M. Rowan College of Engineering,

Rowan University, 600 North Campus Drive, Glassboro, NJ08028, USA

(Tel: 856-256-5375; e-mail: yenkie@rowan.edu; ghanem@students.rowan.edu)

**Abstract:**

Chromates are widely used for their anticorrosive properties. Unfortunately, they are highly hazardous with environmental agencies regulating their levels to below 10 ppb in drinking water. As anion exchange resins are typically used for removal, predictive dynamic models are necessary to make quick decisions rather than relying on experimental data that could take several days to implement. With various dynamic models currently applied to simulate the ion exchange process, the Thomas model was picked for its simplicity and better accuracy when compared to other models. The Thomas model contains two parameters, the constant ($K_T$) and the maximum resin capacity ($q_m$), which are empirically calculated. Unfortunately, the model demonstrated large parameter fluctuations with no correlation to varying contact times or inlet chromate concentrations. Therefore, fixing both parameters will lead to failed model predictive behavior. By fixing the value of $q_m$ and proposing a linear relationship of $K_T$ with resin contact time and inlet chromate concentration, the accuracy of the model was improved five-fold, demonstrating its potential for better process controls.

**Statement of Industrial Relevance:** Since anion exchangers are usually applied as a single use resin bed to remove chromates to ppb levels, decision on flow rate through the resin bed would need to be made based on the inlet concentration rather than the outlet concentration. Therefore, a more robust model with parameters that are relatable to inlet chromate concentration and flow rate changes is needed to allow for quick decisions and to maximize the performance of the resin bed.

**Novelty or Significance:** Most predictive models used today have parameters that need to be empirically calculated at each ion exchange resin system condition such as flow rate and inlet concentration. But due to the low inlet concentrations, these empirical models take 100s of hours of experimental work to develop a relationship at each system conditions based on the breakthrough capacity. Therefore, a model with fixed parameters and/or parameters with direct relationships to the system conditions would be a significant improvement over the current models allowing for quick decisions before a premature breakthrough of toxic compounds into the water supply.

**Introduction:**

Chromates, while necessary in industrial applications, are highly hazardous if allowed to get into the drinking water supply, even at low ppb levels. Therefore, accurate predictive models are needed to achieve the proper removal of chromates. Unfortunately, most models have empirically calculated parameters that are too sensitive to fluctuations of the inlet chromate concentrations or flow rates. In this report, we propose a modification of the model to improve the model predictability therefore allowing for better control strategies and avoiding early breakthrough of the toxic compound.

**Chromate History and Regulations:** Chromates are oxyanions, with the generic formula $A_xO_y^{2-}$, where A represents a chemical element and O represents an oxygen atom, typically found in the VI oxidation state, toxic due to its high oxidizing power. Chromates have been widely used for many years as corrosion inhibitors or as alloys for metal adhesion in the steel, aluminum, zinc and other metal industries [1]. In addition to its decorative nature in the automotive sector, chromates possess good electrical conductivity when used in the semiconductor industry and is used in redox chemical reactions as titrants in analytical laboratories[2].

In 2002, the European Union initiated several policies restricting the use of hexavalent chromium and controlling the spread of the genotoxic carcinogen responsible for mutagenic damage[3]. The strong oxidative effect of chromium breaks the DNA[4] since the cell walls will allow the oxyanion to be absorbed intracellularly via the sulfate channels due to their structural similarities. Based on these studies, a maximum limit of 100 ppb was imposed by the European Union[5]. In the United States, an increased incidence of lung cancer was observed among employees who were subjected to chromates via chronic inhalation whereas an increase in liver, stomach, and intestine cancer occurred via drinking water with relatively unsafe levels of chromium[6]. In 2012 and based on these findings, the OSHA and EPA imposed limits of 5 ppb for airborne exposure and 10 ppb for drinking water.

**Removal of Chromates from Drinking Water:** There are many ways to control the levels of chromium within the drinking water such as reducing the chromium 6+ to the less toxic chromium 3+[7] or removing it via purification methods such as activated carbon[8], precipitation followed by filtration[9], or ion exchange. The latter solution of using ion exchangers has grown significantly since it is capable of achieving parts per billion (ppb) level removal [10]. As seen in Figure 1, chromates are mainly available as negatively charged oxyanions at drinking water pH range (from 6.0 to 8.5) with mainly $CrO_4^{2-}$ and $Cr_2O_7^{2-}$ being the most commonly found ions[11]. For such divalent negative ions, the resin of choice would be the positively charged ion exchange resin, generally known as an anion exchanger[12].

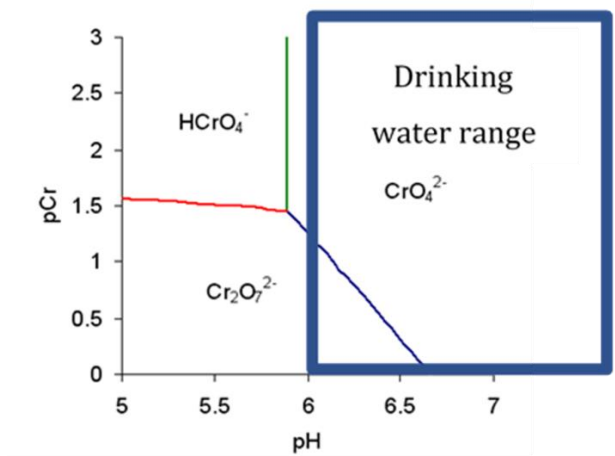

Figure 1: Predominance diagram for chromates showing the commonly found oxyanions based on the pH of the water and the logarithmic concentration of the chromate displayed as pCr = -log[Cr].

One of the most common anion exchange resins is the standard microporous with trimethyl amine functionality, making it a strongly basic quaternary amine anion exchanger. Such resin is widely used in demineralization applications[13] as it removes all types of anions that exist in the solution such as chlorides, sulfates, bicarbonates, etc. Therefore, the anion exchanger is expected to work well in removing the negatively charged chromates. In this paper, an anion exchange resin, Purolite A600E[14], was used to find a practical model to predict the performance for chromate removal and determine the key variables and parameters affecting the overall process. Purolite A600E will also be compared to another anion exchange resin, Purolite A520E[15], that is porous (as opposed to microporous) with triethyl amine (as opposed to trimethyl amine) functionality to confirm if such modeling can predict the performance of different anion exchangers. The bulkier ligand of A520E makes it less selective for divalent anions like chromates. The Purolite A600E and A520E data sheets are provided in the supplementary section. An understanding of chromate extraction will be investigated via systematic modeling and simulation.

**Modeling Approaches:** As the goal is to remove chromates down to low ppb levels, a continuous ion exchange or fixed bed adsorption process would be required since a batch operation can only achieve equilibrium values, not enough to achieve the required low limits. There are several computational models to predict the dynamics of a continuous process; such as the widely used Thomas model [16–19](reported in many publications as the Bohart-Adam model[18,20]) in a fixed bed system. All these models possess empirical parameters that are calculated via minimization of the error between the experimental and predicted values[18,21]. Unfortunately, such parameters are not fixed as they depend on many system variables such as inlet concentrations and flow rates of the system. The goal of this report is to modify the relationship needed between the model parameters and the system characteristics to improve the model predictability, therefore allowing process control decisions to be made. Since the ion exchange resin will not be reused after being loaded with the toxic chromate, the improved model will help to maximize the loading based on the inlet chromate concentration by manually adjusting the inlet flow rates.

**Methodology:**

**Data Collection:** The data collected was extracted from the work of Xue Li and coworkers from the University of California at Davis[22–24]. The experiments consisted of periodical measurement of the effluent chromate concentration when a specific inlet chromate contaminated water feed is introduced. In their work, summarized in Table 1, they started by packing two different resin quantities of Purolite A600E, a standard microporous anion exchanger with quaternary amine functionality, in different size columns varying the flow rates of a 14.73 ppb inlet chromate feed, keeping the resin contact time constant, as demonstrated in experiment 1 and 2. It was followed by running the same feed at different contact times, with the flow rate and linear velocity constant, seen in experiments 1 and 3.

The work was repeated by using a different feed source with a higher inlet chromate concentration of 44.47 ppb. The contact time was changed from 0.5 min to 0.75 min as described in experiments 4 and 5 seen in Table 1. A final experiment 6 with a third water source contaminated with chromate at 20.65 ppb concentration was used, as noted in Table 1. This final feed would confirm the accuracy of the model built from the results of the first 5 experiments.

The water feed from experiment 1 was also used for a different resin, A520E, using the conditions for experiment 1 (0.75 min) and for experiment 3 (0.5 min). The model will be calculated for this resin as well to confirm the usefulness of such model independent of which resin being used.

The columns were continuously fed with the chromate contaminated water until either complete exhaustion, where the effluent concentration reached the influent chromium concentration, or where the breakthrough point was reached, defined as the maximum allowed chromate leakage or 10 ppb. The experimental data and a more detailed method description are summarized in the supplementary information.

Variations of the inlet chromate concentrations ($C_0$), flow rates (Q), linear velocities, contact time, resin volume, and the column diameter are all summarized in Table 1.

Table 1: Summary of the experimental studies done by Xue Li and coworkers with Purolite A600E (experiment 1 to 7) and with Purolite A520E (experiment 8 and 9) to remove chromates[22–24]

| Experiment Number - Resin | $C_0$ ppb Cr | Q-flow Rate l/hr | Linear Velocity cm/min | Contact Time min | Resin Volume ml | Column Diameter cm |
|---|---|---|---|---|---|---|
| **1-A600E** | 14.73 | 0.85 | 8 | 0.75 | 10.6 | 1.5 |
| **2-A600E** | 14.73 | 457 | 22 | 0.75 | 5712 | 21 |
| **3-A600E** | 14.73 | 0.85 | 8 | 0.5 | 7.1 | 1.5 |
| **4-A600E** | 44.47 | 0.85 | 8 | 0.75 | 10.6 | 1.5 |
| **5-A600E** | 44.47 | 0.85 | 8 | 0.5 | 7.1 | 1.5 |
| **6-A600E** | 20.65 | 40.80 | 21 | 0.75 | 510 | 6.4 |
| **7-A520E** | 14.73 | 0.85 | 8 | 0.75 | 10.6 | 1.5 |
| **8-A520E** | 14.73 | 0.85 | 8 | 1.5 | 7.1 | 1.5 |

The concentrations of other competing ions such as chlorides, sulfates, nitrates, and bicarbonates were also reported for these experiments in the supplementary section. But as the levels of the competing ions are already hundreds of times in excess of chromates, their variations are assumed to have minimal effect on the performance of chromate removal [25].

**Dynamic Model Discrimination:** One of the most reported dynamic models is the Thomas model shown in its linearized standard form in Eq 1.

$$\ln\left(\frac{C_0}{C(t)} - 1\right) = K_T q_m \frac{M}{Q} - K_T C_0 t \qquad (Eq1)$$

where $C_0$ - inlet concentration, C(t) - effluent concentration varying with time, $K_T$ - Thomas constant (model parameter), $q_m$ - maximum resin capacity (model parameter), M - mass of resin, Q - flow rate through the system, and t – time. Other widely applied dynamic column models are the Clark model that depends on the Freundlich batch process to simulate a continuous process[20], the Yoon-Nelson model[18], and the Wolborska model which deals with low concentration breakthroughs[20]. These models are summarized in Table 2. The parameters in these models can be calculated by minimizing the difference between the experimental data and the model-predicted values using a nonlinear optimization program [26].

In this work, the Thomas model was picked for its simplicity, its relationship with some system characteristics, and for fitting the experimental data more accurately than the other models. The advantages and disadvantages of the four models are summarized in Table 2 and the graphical presentation can be seen in Figure 2.

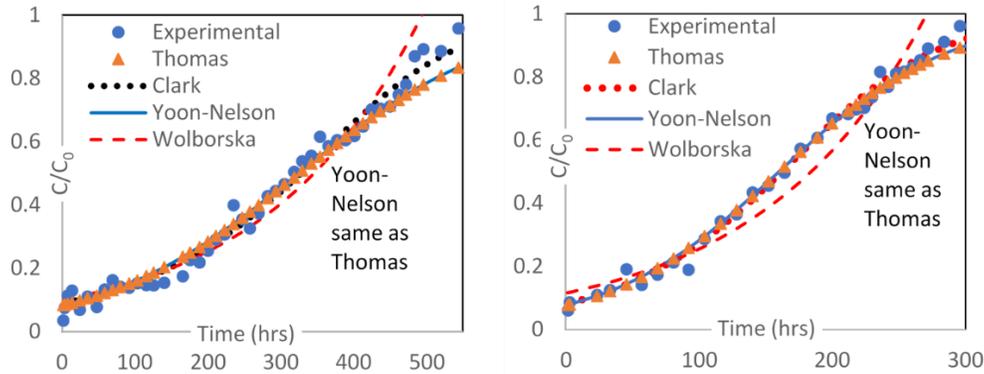

Figure 2: Comparison of the various kinetic models for chromate removal with A600E (left) and with A520E (right). Both graphs show Wolborska's accuracy limitation to low chromate. The other 3 models offer better fit while the Clark model needs 3 parameters in its prediction making it too difficult to simplify.

Table 2: Summary of models considered for the chromate removal work

|  | Thomas[17,19,27] | Yoon-Nelson[16] | Clark[28] | Wolborska[28] |
|---|---|---|---|---|
| *Equation* | $\ln\left(\frac{C_0}{C} - 1\right) = \frac{K_T q_m M}{Q} - K_T C_0 t$ | $\frac{C}{C_0} = \frac{\exp(K_{YN} t - \tau K_{YN})}{1 + \exp(K_{YN} t - \tau K_{YN})}$ | $\left(\frac{C_0}{C}\right)^{n-1} - 1 = A e^{-rt}$ | $\ln\left(\frac{C}{C_0}\right) = \frac{\beta_a C_0}{N_0} t - \frac{\beta_a Z}{U_0}$ |
| *Parameters* | $K_T, q_m$ | $K_{YN}, \tau$ | $A, r, n$ | $\beta_a, N_0$ |
| *Variables* | $\left(\frac{C}{C_0}\right), t$ | $\left(\frac{C}{C_0}\right), t$ | $\left(\frac{C}{C_0}\right), t$ | $\left(\frac{C}{C_0}\right), t$ |
| *System Characteristics* | $M, Q, C_0$ | none | none | $U_0, C_0, Z$ |
| *Attributes* | related to the most common isotherm model = Langmuir | Simplest form with 2 variables and 2 parameters | Related to the oldest batch isotherm model = Freundlich | Describes breakthrough at low concentrations |
| *Advantages* | Can be linearized + related to 3 system characteristics | Can be linearized | 3 parameters for better accuracy | Good fit at low concentrations + Can be linearized |
| *Disadvantages* | Computed $q_m$ is different than experimental maximum resin capacity | No relations to system variations. Simple representation of the Thomas model with $K_{YN} = K_T C_0$ and $\tau = \frac{q_m M}{C_0 Q}$ | Cannot be linearized with $n$ usually taken from batch isotherm experiments. | Limited to film diffusion kinetics. $\beta_a$, is a function of two parameters and one system constant [28] |

**Model Adjustment:** As mentioned previously, the Thomas model was picked for its simple representation of the relationship with the total resin mass in the column, the system flow rate, and the inlet chromate concentration. But further adjustments are needed with the Thomas model to reflect the measured concentration ratio, $\frac{C}{C_0}$, versus time.

Eq 1 can be rewritten into a nonlinear exponential expression as follows:

$$\frac{C}{C_0} = \frac{1}{\exp\left(K_T q_m \frac{M}{Q} - K_T C_0 t\right)+1} \quad (\text{Eq 2})$$

$\frac{C}{C_0}$ is preferred over the Thomas model display of $\frac{C_0}{C}$ to avoid a large range of ratio variations due to the effluent concentration, C, being close to zero at the start of the process. $\frac{C}{C_0}$ can only vary between 0 and 1 allowing for better computational accuracy. When using $\frac{C}{C_0}$ in Eq 2, one can calculate the time needed to achieve 10% breakthrough ($\frac{C}{C_0} = 0.1$) or 50% breakthrough ($\frac{C}{C_0} = 0.5$), related to the critical half-time concentration value (by zeroing the exponential number in Eq 2).

Since ion exchange resins are typically measured by volume as opposed to mass as done with activated carbon, the Thomas model in Eq 2 was adjusted further to reflect resin volume, $V$, rather than mass, $M$. This will mainly impact the units of the maximum resin capacity, $q_m$, as it will change from grams of Cr per kilogram of resin to grams of Cr per liter of resin as the resin density is assumed constant during use. Therefore, the final form of the Thomas model is seen in the non-linear representation below:

$$\frac{C}{C_0} = \frac{1}{\exp(K_T q_m\, CT - K_T C_0 t)+1} \quad (\text{Eq 3})$$

where CT – $Contact\ time\ =\ \frac{V}{Q}$ (also called residence time)

**Calculations of the parameters:** The parameters in the Thomas model can be identified by minimizing the difference between the experimental data and the model simulations using nonlinear optimization where the objective function to be minimized is the Residual Sum of Squares Error between model predictions and experimental measurements[17,18,29–32]. The RSSE form listed below is used to normalize the error being minimized:

$$RSSE\ =\ Residual\ Sum\ of\ Squares\ of\ Error\ =\ \sum_i \left(\frac{Y_{ical}-Y_{iexp}}{Y_{ical}}\right)^2 \qquad \text{(Eq 4)}$$

Where $Y_{iexp}$ - ith data of the experimental measurement $\left(\frac{C}{C_0}\right)_{iexp}$, and $Y_{ical}$ - ith predicted or calculated value $\left(\frac{C}{C_0}\right)_{ical}$.

The hybrid error function[31,33] is also calculated to confirm the accuracy of the model.

$$Hybrid\ fractional\ error\ function\ =\ HFE\ =\ \frac{100}{n-p}\sum_i \frac{(Y_{ical}-Y_{iexp})^2}{Y_{ical}} \qquad \text{(Eq 5)}$$

Where n - number of data points (varies between the different experiments) and p - number of parameters = 2 (for $q_m$ and $K_T$).

**Sensitivity analysis of the results:** With every empirical equation, sensitivity analysis is used to confirm what parameter deviation will have the most impact on the model prediction accuracy. The impact of the two computed parameters, $q_m$ and $K_T$, on the normalized chromate concentration, $\frac{C}{C_0}$, is checked by evaluating the derivatives of the concentration function of the model in respect to the parameters [34]. A generalized partial differential equation (6) is seen below:

$$P(Y\ p_i) = \frac{\delta f(K_T, q_m, C_0, CT;\ t)}{\delta p_i} \tag{Eq 6}$$

where f () - function of time representing the Thomas model in Eq 3, $p_i$ - ith parameter or variable such as $K_T$, $q_m$, $C_0$, $CT$, and $P$ - partial derivative of the Thomas equation $Y = f(K_T, q_m, C_0, CT;\ t)$ in respect to $p_i$.

Therefore, two sensitivity equations can be written based on Eq 6 for variations with the computed parameters.

$$P(Y\ K_T) = \frac{\delta Y}{\delta K_T} = \frac{(C_0 t - q_m CT)\exp(K_T q_m CT - K_T C_0 t)}{(1+\exp(K_T q_m CT - K_T C_0 t))^2} \tag{Eq 7}$$

$$P(Y\ q_m) = \frac{\delta Y}{\delta q_m} = -\frac{K_T\ CT\ \exp(K_T\ q_m\ CT - K_T\ C_0\ t)}{(1+\exp(K_T\ q_m\ CT - K_T\ C_0\ t))^2} \tag{Eq 8}$$

Eq 7 shows that function deviation with $K_T$ is more dependent on time (t) than with $q_m$.

**Results and Discussion:**

The results' section is divided into 4 main subsections with the first part looking into the effect of the parameters where the column dimensions are changed between experiment 1 and 2 in Table 3. The second part consists in studying the effect of varying the contact time on the parameters as demonstrated by experiment 3. This is followed by evaluating the effect of increasing the inlet chromate concentration to 44.47 ppb as seen in experiments 4 and 5. The final subsection develops a workable relationship between the parameters and the operating conditions to confirm, with a final experiment using a different chromate feed, the resulting improvement of the predictive model.

As mentioned in the methodology section 2.3, the parameters are calculated via the minimization procedure for each experiment and are summarized in Table 3. It is observed that the parameters are not fixed as they vary considerably between each experiment.

Table 3: Summary of the results obtained by minimizing RSSE for each experiment

| Experiment Number | $C_0$ ppb Cr | Contact Time min | Linear Velocity cm/min | $K_T$ l/gram Cr.hr | $q_m$ gram Cr/l |
|---|---|---|---|---|---|
| 1 | 14.73 | 0.75 | 8 | 502 | 0.3828 |
| 2 | 14.73 | 0.75 | 22 | 434 | 0.3964 |
| 3 | 14.73 | 0.5 | 8 | 1264 | 0.2549 |
| 4 | 44.47 | 0.75 | 8 | 1612 | 0.2091 |
| 5 | 44.47 | 0.5 | 8 | 2548 | 0.1687 |

With no clear correlations between the parameters and the changing operating conditions, the accuracy of the Thomas model, by fixing both parameters or fixing one parameter while varying the other one, is evaluated.

**Effect of changing column dimensions on the Thomas model:** As described in the methodology section, the first part of the experiments (1 and 2) is to confirm the Thomas model sensitivity to changes in the column dimensions when both inlet chromate concentration and contact time are kept constant. The parameters from the optimized fit of Experiment 1 are used to compute the data in experiment 2 and are plotted in Figure 3B.

From Figure 3A, the residual sum of squares (RSSE) and the regression coefficient ($R^2$) are calculated as 0.082 and 0.975 confirming that the parameters obtained from minimizing the RSSE are accurate. The resulting parameters from experiment 1, $K_T$ and $q_m$, are found to be 502 l/g Cr-hr and 0.3828 g Cr/l respectively. When these parameters are used to predict the results in experiment 2 where the volume of the resin is increased from 10.6 ml to 5712 ml keeping the contact time and inlet chromate concentration constant, the RSSE increased to 0.12 which is an acceptable increase and shown in Figure 3B. The hybrid function error, HFE, also increased from 0.64% in experiment 1 to 2.44% in experiment 2. A confidence band, graphed in Figure 2C, confirms that all the experimental data from experiments 1 and 2 seem to fall within the 5% variance.

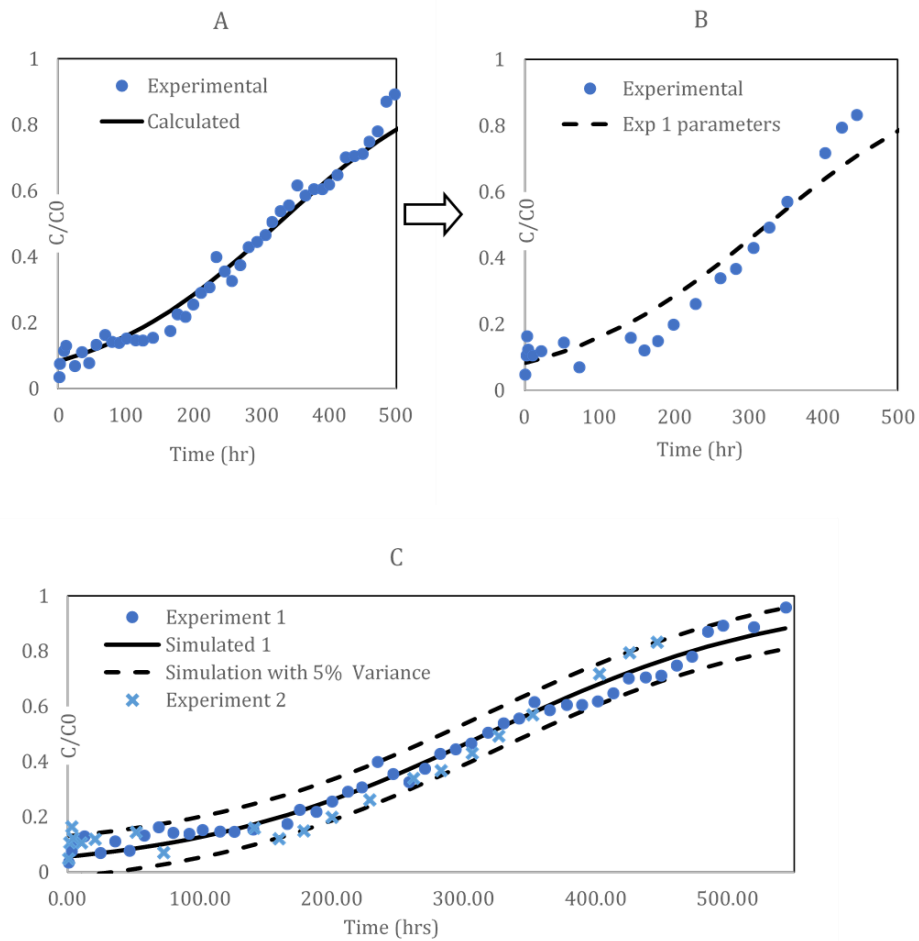

Figure 3: Fitting the model obtained from experiment 1 into experiment 2. A) Fitting model using non-linear regression in the lab experiment 1 using 10.6 ml of A600 at 0.75 min contact time and 14.73 ppb inlet chromate concentration. B) Using model from experiment 1 to predict the data for a larger size column into experiment 2 using 5712 ml of A600 keeping contact time and inlet chromate concentration constant. C) Figure showing that the lab size experiment and the scale-up fall within the 5% confidence band around the predicted model.

Therefore, the system has low sensitivity to the changes in the column dimensions, the flow rates, and the volume of the resin used as long as the inlet chromate concentration and the resin contact time are constant. As a consequence, fixing both parameters will provide good predictive model behavior when scaling up operations without varying the feed concentration or contact time.

**Effect of changing the contact time at constant inlet chromate concentration:** From Figure 4, the contact time of 0.75 min in experiment 1 decreased to 0.5 min in experiment 3. The RSSE and the HFE increased significantly to 0.94 and 11.72% respectively (from 0.082 and 0.64%) when the contact time is decreased by 33% between the experiments. This proves that the Thomas model is highly sensitive to contact times and therefore the parameters, $K_T$ and/or $q_m$, are expected to be dependent on the contact time deviations.

This sensitivity is likely due to the lack of the Thomas model to confirm if the chromate removal is surface reaction limited or diffusion limited. With longer contact time, more diffusion into the resin pores occurs, therefore becoming ion exchange reaction limited. But shorter contact time overwhelms the limitation of the pore diffusion kinetics therefore causing a faster increase in the effluent chromate levels as seen in Figure 4 (diffusion limited). As a result, the Thomas model shows high sensitivity towards contact time deviation due to the model limitations in considering the resin matrix properties. It is logical to hypothesize that, if the resin doesn't have any pores or has pores that are significantly large (minimizing the effect of pore diffusion), a much lower sensitivity to contact time deviation is expected.

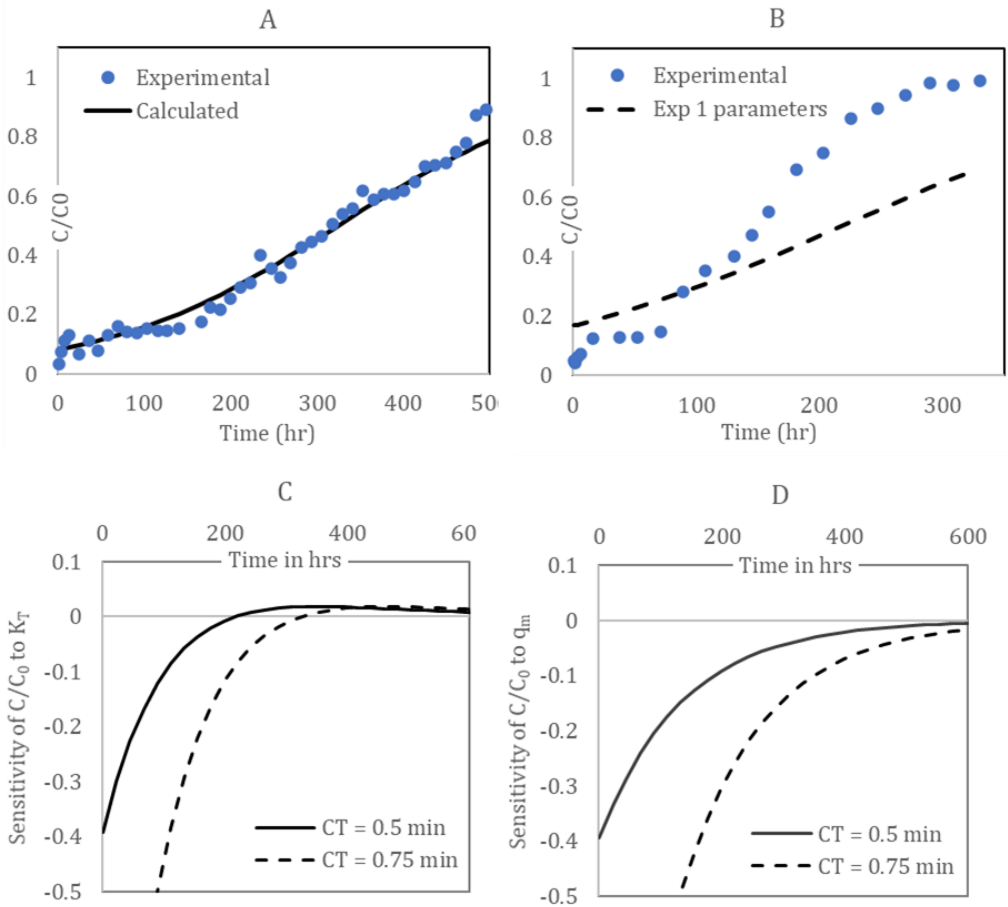

Figure 4: Effect of changing the contact time using the parameters obtained from Experiment 1 to predict the performance in A) Experiment 1 at 0.75 min contact time and B) Experiment 3 at 0.5 min contact time. Both using the same feed with 14.73 ppb chromate concentration. Sensitivity data based on 2 different contact times using the parameters from the fitting models of experiment 1 at constant chromate concentration of 14.73 ppb with C) showing the model sensitivity to 5% variation of the $K_T$ values and D) showing the model sensitivity to 5% deviations of the $q_m$ values.

The sensitivity of the concentration ratio to $K_T$ and $q_m$ variations are plotted for experiment 1 in Figure 4 as well. The data shows, from the rate of change to the deviation of the parameters, that the Thomas model is less sensitive to deviations of $K_T$ and $q_m$ at short contact time allowing the system to reach equilibrium faster.

If boundary conditions are added where $q_m$ and $K_T$ can deviate using the parameters obtained from experiment 1, the parameter estimation via nonlinear programming in Matlab® 'fmincon' is repeated via minimization of the RSSE. The results are seen in Figure 5 using ±30% parameter deviation from the optimized conditions of experiment 1.

Figure 5 demonstrates that some experiments, such as experiments 4 and 5, do not reach the optimized parameters for both $q_m$ and $K_T$ as they fall into the corner of the boundary conditions graph. In such corner positions, acceptable optimized conditions are unachievable when fixing either parameter. On the other hand, experiment 3 is able to achieve optimized conditions on only $q_m$, with $K_T$ being limited by the ±30% range, therefore falling on the maximum line. Such values achieve partial optimization which might deliver acceptable predictions when one of the parameters is fixed. Experiment 2 is the only experiment that is optimized within the imposed boundaries and confirms the conclusions made in section 3.1, that changing the column dimensions has no significant effect on the model prediction using the parameters from experiment 1.

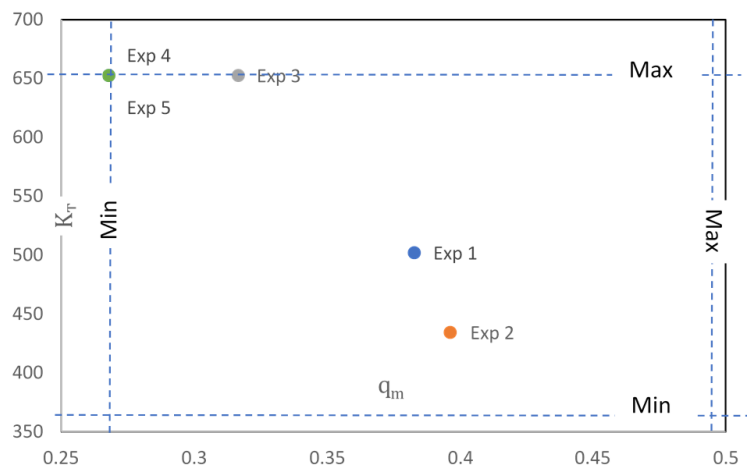

Figure 5: ±30% upper and lower boundary conditions imposed on the parameters based on the optimized values for experiment 1. Therefore, the parameters of defining experiment 1

is at the center of the rectangle with the minimum and maximum lines marking the 30% minimum and maximum boundary conditions while optimizing the other experiments.

**Effect of changing the inlet chromate concentration at constant contact time:** When varying the inlet concentration of the chromate by changing the source of the contaminated water used, we are able to plot such effect in Figure 6. In this experimental work, the Thomas model experienced its largest error deviation. In spite of the Thomas model having the inlet chromate concentration, $C_0$, incorporated in its equation, there is a large deviation of the predicted values from the experimental data when applying the parameters obtained from experiment 1. The HFE increases from 0.64% for experiment 1 to 15.22% in experiment 4.

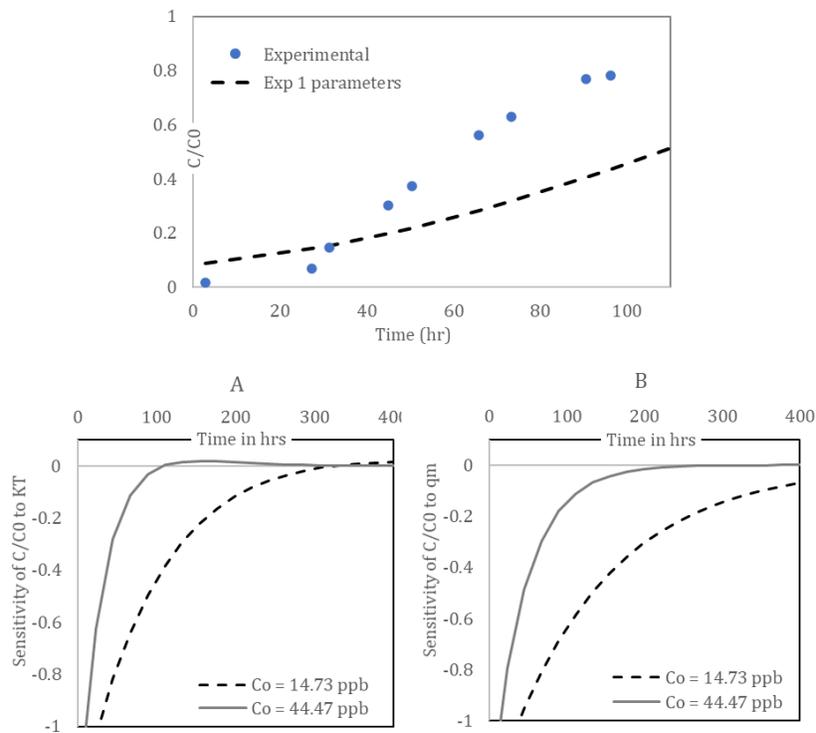

Figure 6: Top graph shows the effect of changing contact time using the parameters from Experiment 1 to predict the performance in experiment 4 at 0.75 min contact time proving that it is a bad fit. Graph A and B show the sensitivity data based on 2 different chromate concentrations using the parameters from the optimized model of experiment 1 at constant

contact time of 0.75 min with A) showing the model sensitivity to 5% variation of the $K_T$ values and B) showing the model sensitivity to 5% deviations of the q_m values.

This prediction failure, as mentioned in section 3.2, can also be attributed to ignoring the resin characteristics such as pore diffusion and surface reaction limitations affecting the efficient removal of chromate. The sensitivity of the inlet chromate concentration on $K_T$, and $q_m$ are also plotted in Figure 6. The sensitivity from the parameters' variations is more prevalent with low chromate concentration than high chromate concentration. It is thought that high chromate concentration difference, between the solution and the internal resin matrix, improves the diffusion with increased osmosis therefore minimizing such variations.

Since the sensitivity seems to be less prevalent in higher concentration and, if we add some boundary conditions allowing $q_m$ and $K_T$ to deviate from the parameters obtained from experiment 4, we can run the non-linear optimization work that gave us the results in Table 3 via minimization of the RSSE. The results are shown in Figure 7 using ±30% parameter deviation from experiment 4 optimized conditions.

Figure 7 shows that there is no experiment with the optimized parameters for both $q_m$ and $K_T$ falling into the corner of the boundary conditions graph as seen in the previous section. Other experiments, such as experiments 1, 2, and 5, are able to achieve optimized conditions on only 1 parameter, mainly $q_m$. Experiment 3 is the only experiment that is optimized within the imposed boundaries. This is an improvement over the previous results in section 3.2 confirming that using higher concentration of chromate will allow for less sensitivity to deviations in the parameters.

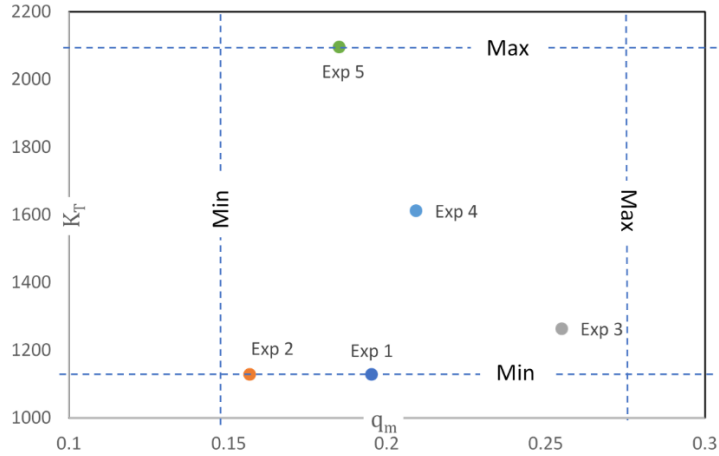

Figure 7: ±30% upper and lower boundary conditions imposed on the parameters based on the optimized values for experiment 4.

**Revising the Thomas model parameters to improve accuracy and confirmation of the results:** As the boundary conditions of ±30% around the parameters for experiment 1, at low chromate concentration, and for experiment 4, at high chromate concentration, fail to give acceptable results for the all the experiments, the average value of the parameters for experiment 1 to 5 are used to replot the graphs displayed before in Figure 4 and 6. The parameters for experiment 2 are not used in the average since it has already been determined that scaling up the purification step with the same inlet chromate concentration and contact time lead to similar predictions. The average values of $q_m$ and $K_T$ are calculated as 0.254 gram/l-resin and 1357 l/gram-hr respectively. Same as in sections 3.2 and 3.3 with boundary conditions where we allow $q_m$ and $K_T$ to deviate from the average parameters calculated above, the non-linear optimization is run using ±30% parameter deviation that results in Figure 8.

Figure 8 shows that there is no experiment that has both parameters on the corners of the accuracy rectangle. Other experiments, such as experiment 1, 2, and 5, are able to achieve optimized conditions on only 1 parameter, $q_m$, with only $K_T$ limited by the boundary conditions. Experiment 3 and 4 are optimized within the imposed boundaries. While no $q_m$ values for any experiment fall on the 30% boundary line, only the $K_T$ values of some experiments seem to reach the 30% boundary conditions.

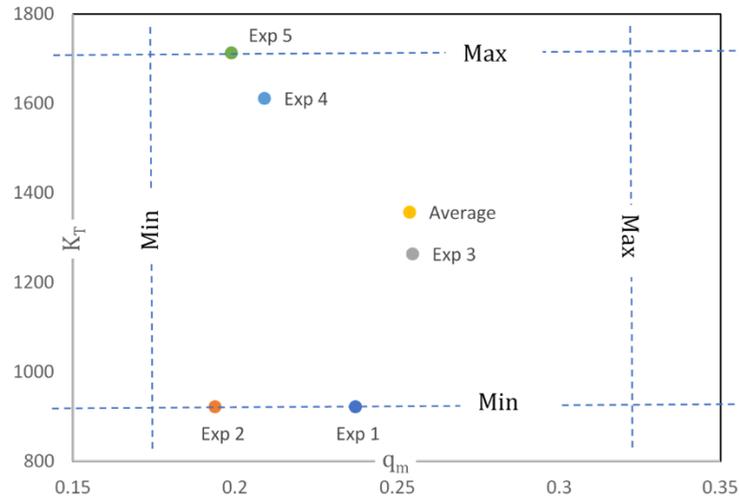

Figure 8: ±30% upper and lower boundary conditions imposed on the parameters based on the average parameter values of experiment 1, 3, 4, and 5 that are shown in Table 3.

Based on these results, the following assumptions were made:

1- Parameters $q_m$ and $K_T$ are considered independent from each other and only rely on changes in the system.

2- Since the parameter $q_m$ is related to the maximum resin capacity for a particular ion removed and since all $q_m$ values fall within the ±30% range, the value of $q_m$ was fixed allowing only $K_T$ to vary. The value taken for $q_m$ is the mean value of all the $q_m$ calculated from the experiments 1 to 5 in Table 3. The mean value is calculated at 0.254 gram Cr/l.

3- With $K_T$ as the remaining parameter allowed to deviate with changing contact times and inlet chromate concentrations, the values of the calculated $K_T$ from Table 3 are recalculated by optimizing the nonlinear programming, fixing one parameter $q_m$ and allowing $K_T$ to vary until optimization is achieved by minimizing the RSSE. The optimized values are seen in Table 4 and graphed in Figure 9. A simple linear regression is assumed between $K_T$ and contact time as well as $K_T$ and inlet chromate concentration.

Table 4: Summary of the results obtained by minimizing RSSE for each experiment fixing the $q_m$ parameter at 0.254 grams Cr/l

|  | Values for the experimental parameter $K_T$ with fixed $q_m$ | | | | |
| --- | --- | --- | --- | --- | --- |
|  | #1 | #2 | #3 | #4 | #5 |
| $K_T$ (l/gram.hr) | 769 | 653 | 1269 | 1080 | 1146 |

From the two relationships plotted in Figure 9, a single 3-dimensional linear relationship can be extrapolated as seen below.

$$K_T = -264\ CT + 10.45\ C_0 + 1247 \tag{Eq 9}$$

With CT having units of minutes and $C_0$ having units of ppb.

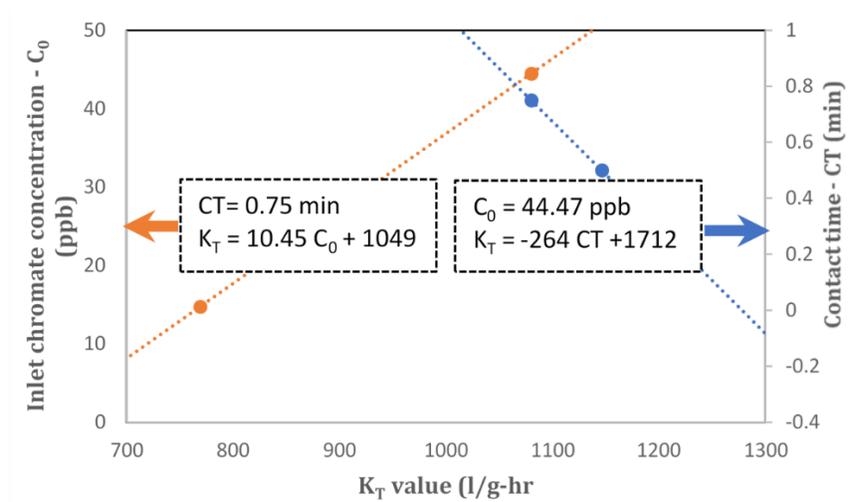

Figure 9: Parameter $K_T$ deviation with changes in system characteristics such as varying inlet chromate concentration at constant contact time of 0.75 min and varying contact time at constant inlet chromate concentration of 44.47 ppb.

With the new relationships introduced, experiment 6 is performed to confirm that a model improvement is achieved by allowing $K_T$ to change with contact time and chromate concentration. In experiment 6, an inlet chromate concentration of 20.65 ppb and a contact time of 0.75 min using a larger resin volume, as described in Table 2. The predicted model for experiment 6 using the parameters from experiment 1 is compared with the new predicted model using the proposed parameters and is plotted in Figure 10.

The following parameters for experiment 6 are obtained: The fixed $q_m$ = 0.254 gram Cr/l and the adjusted $K_T$ = 1265 l/gram Cr-hr. Figure 10 shows that, allowing one of the parameters to change with contact time and inlet chromate concentration, the new model has improved the predicted values since the hybrid function error, HFE, calculated at 6.3%, is a five-fold decrease from the 32.7% using the fixed parameters from experiment 1.

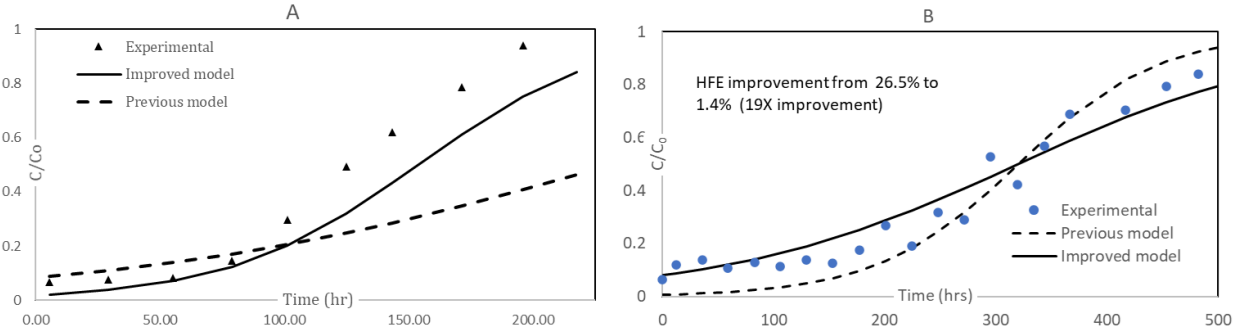

Figure 12: A) A600E - Comparison of the previous model using experiment 1 parameters with the new model allowing $K_T$ to change with contact time and inlet chromate concentration. Both models were compared in experiment 6 using 20.65 ppb inlet chromate concentration and 0.75 min contact time. B) A520E - Comparison of the previous model using experiment 7 parameters with the new model allowing $K_T$ to change with contact time. Both models were compared in experiment 8 using 14.73 ppb inlet chromate concentration and 0.5 min contact time.

When the work was repeated with the other resin, A502E, by fixing its $q_m$ at 0.1886 g/L (averaging $q_m$ and linearizing the changes of the $K_T$ with the system characteristics, calculated as $K_T = -724.4\ CT + 1603.3$, from experiments 7 and 8), Figure 10 reconfirms the previous result with A600E and shows improvement in the accuracy of the proposed method (19X reduction in HFE) when using A520E.

**Conclusions:**

In this work, the Thomas model has been used to predict the performance of chromate removal from drinking water applications using A600, a microporous strong anion exchanger resin from Purolite. The Thomas model has been picked since it considers other system characteristics such as flow rates, resin volume, and inlet concentration and it shows better predictability than other models. By minimizing the residual sum of square error between the experimental and calculated concentration ratio, the two parameters $K_T$ and $q_m$ have been determined. After modeling 5 experiments varying the contact times and chromate concentrations, only changes in column dimensions and resin volume, when contact time and chromate concentration are fixed, were predicted successfully by the Thomas model. On the other hand, variations in contact time and inlet chromate concentration have led to a dramatic failure in the prediction accuracy of the outlet concentration of the system.

An improved model was proposed in this work by fixing one parameter, $q_m$, using an average value of several experiments, and by allowing the other parameter, $K_T$, to change with contact time and inlet chromate concentrations following a simple linear relationship. The improved model has led to a 5-fold decrease of the hybrid fractional error function. Without the need to run more empirical work which could take many days, such improvement in the model accuracy will help to extend the resin life by manually adjusting the flow rate when the inlet chromate concentrations fluctuate. Figure 14 shows the expected performance if the inlet chromate concentration shifts from 14.73 ppb to 20.65 ppb without adjusting the initial contact time of 0.5 min versus a quick adjustment to 0.75 min.

For future work, we will extend the model to allow for time dependent optimal control strategy so the flow rate can be automatically adjusted based on the chromate concentration entering the anion exchange column. Extending the model to the controlling parameters will allow the minimization of the resin quantity for maximum processing time.

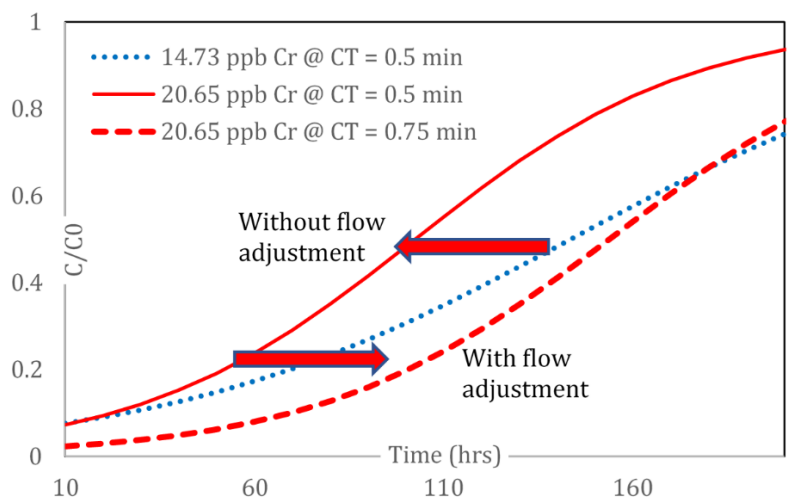

Figure 14: Expected performance of the A600E when using the improved model at 14.73 ppb inlet chromate concentration and 0.5 min contact time. The resin will fail earlier (shifts to the left) when the inlet chromate concentration increases to 20.65 ppb. If the flow was adjusted immediately when the inlet chromate concentration increased, the performance of the ion exchange resin will improve as more chromate loading can be achieved (graph shifts to the right)